\documentstyle[psfig,epsf,amsmath]{article}
\begin{document}

\begin{center}
{\Large \bf Bootstrap Equations for String-Like Amplitude} \\

\vspace{4mm}

Kirill Semenov-Tian-Shansky\\
St. Petersburg State University,
Universit\'e  de Li\`ege au Sart Tilman \\
 E-mail: semenov@pdmi.ras.ru\\
\end{center}

\begin{abstract}
One of the ways to check the consistency of our effective field theory
(EFT) approach
(see \cite{qfthep})
is to perform the numerical testing of those sum rules for hadron
resonance parameters which follow from the system of bootstrap
constrains. In this talk we discuss the peculiar features of this
procedure for the case of exactly solvable bootstrap model based on
Veneziano string amplitude. This allows us to simulate different
situations that may encounter in realistic EFT models. We also make a
short review of the technique that may be useful for further
analysis of various bootstrap systems.
\end{abstract}


\section{Introduction}
\mbox

In papers
\cite{AVVV}-\cite{Essential}
an attempt is made to develop an effective field theory formalism
suitable for description of hadronic scattering processes at low
energies (see also
\cite{qfthep} and ref. therein).
It was shown that the natural requirements of self-consistency of the
perturbation series for scattering amplitudes lead to an infinite
system of restrictions (called
{\em bootstrap constraints})
for the physical parameters of a theory. Once solved the bootstrap
system would permit to uncover the set of truly independent physical
parameters of the effective field theory. Unfortunately, for the
present we are unable to point out an explicit solution to the
bootstrap system. So, roughly speaking, the only way to check the
consistency of our effective theory approach is to perform the
numerical testing of the relations (sum rules) for physical
parameters of the theory that follow from the bootstrap system. In
other words, we need to check if it is possible to approximately
saturate those sum rules with the finite number of experimentally
known resonances. This kind of verification was done for the cases of
$\pi K, \, \pi N$
and
$KN$
systems (see ref. in \cite{qfthep}).

In each case it proved to be possible to single out certain groups of
sum rules that are very well saturated with known data on the
resonance spectrum. At the same time, there were found sum rules which
cannot be satisfactorily saturated with the same set of data. Clearly,
one of the reasons is that the modern information on hadron spectrum
is far from being exhaustive (especially in the region
$M > 2$ GeV).
That is why only those sum rules that converge sufficiently rapidly
could undergo the experimental verification and posses certain
predictive power. So it would be extremely instructive to learn how to
pick out such sum rules from the infinite system of bootstrap
constraints.

The problem is that even the simplest (tree level) bootstrap system
for the physical parameters of
$2 \rightarrow 2$
hadron scattering looks very intricate and awkward. Its analysis is
complicated by multiple technical obstacles such as the non-trivial
geometry of Mandelstam complex plane, the presence of various
symmetries and bulky expressions corresponding to contributions of the
high spin resonances. That is why it seems reasonable to simulate
different situations that may be encountered in realistic numerical
analysis. For this we need to consider a simple model example. A good
candidate for such tests is the
``toy bootstrap model''
(discussed previously in
\cite{AVVVKS}).
Surely, the main reason to choose it is that this model is exactly
solvable, in a sense that in this case we have in hands the closed
analytical expression for the amplitude that satisfies the system of
bootstrap constraints.

In  this talk we consider the numerical test of sum rules for this
bootstrap model and discuss the methods of accelerating their
convergency. We will also introduce the matrix methods that (as show
our preliminary investigations) may be useful for the further analysis
of bootstrap systems.

\section{Veneziano String Amplitude: Cauchy Forms}
\mbox

We consider a simple model for the scattering amplitude that is
constructed --- in accordance with the idea of Veneziano
\cite{Veneziano} --
out of
$B$-function
without a tachyon:
\begin{equation}
 A(x,y) = (-x-y)B    ( \textstyle \frac{1}{2}-x,  \frac{1}{2}-y ) =
 \displaystyle
 \frac{\Gamma(\frac{1}{2}-x)\Gamma(\frac{1}{2}-y)}{\Gamma(-x-y)}.
 \label{Axy}
\end{equation}
In this Section we give a short summary of the properties of this
function.

The string amplitude has the following specific points (hyperplanes)
$(m,n = 0,1,2,...)$:
\begin{itemize}
 \item
  Zero hyperplanes:
  $ x+y=n. $
 \item
  Pole hyperplanes in
  $x$ ($y$
  fixed,
  $x+y \neq m$):
  $ x=\frac{1}{2}+n. $
 \item
  Pole hyperplanes in
  $y$ ($x$
  fixed,
  $x+y \neq m$): $ y=\frac{1}{2}+n. $
 \item
 There are also three series of ambiguity points located at the
 intersections of the zero hyperplanes with the pole hyperplanes
 in any variable.
\end{itemize}

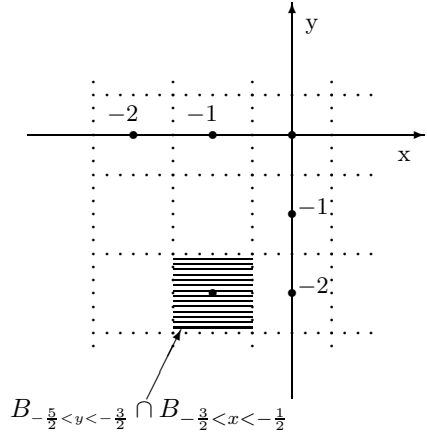
\begin{figure}
\begin{picture}(150,160)
\put(150,110){\vector(1,0){150}}
 \put(250,10){\vector(0,1){150}}
  \put(195,10){\vector(1,2){13}}
\put(140,3){{
$
B_{\scriptscriptstyle -\frac{5}{2}<y<-\frac{3}{2}}
            \cap B_{-\frac{3}{2}<x<-\frac{1}{2}}
$          }}
 \put(290,100){{\small x}}
  \put(255,150){{\small y}}

\put(252,80){{\small $-1$}}
 \put(252,50){{\small $-2$}}

\put(210,115){{\small $-1$}}
 \put(180,115){{\small $-2$}}

\multiput(180,125)(5,0){21}{\circle*{1}}
 \multiput(180,95)(5,0){21}{\circle*{1}}

\multiput(180,65)(5,0){21}{\circle*{1}}
 \multiput(180,35)(5,0){21}{\circle*{1}}

\multiput(175,30)(0,5){21}{\circle*{1}}
 \multiput(205,30)(0,5){21}{\circle*{1}}

\multiput(235,30)(0,5){21}{\circle*{1}}
 \multiput(265,30)(0,5){21}{\circle*{1}}

\multiput(205,37)(0,2){14}{\line(1,0){30}}  

 \put(250,50){\circle*{3}}
  \put(250,80){\circle*{3}}
   \put(250,110){\circle*{3}}

    \put(220,110){\circle*{3}}
     \put(190,110){\circle*{3}}
      \put(220,50){\circle*{3}}

\end{picture}
\caption{\em The plane of intersection of different layers
             $B_x$, $B_y$
             \label{Betabootpic}}
\end{figure}

It can be shown that in the system of layers
$$
B_{n-\frac{1}{2} <y< n+\frac{1}{2}}  \quad n=0,\pm1,...
$$
the amplitude
$A(x,y)$
is the $n$-bounded%
\footnote{In the sense of contour asymptotics; see
\cite{AVVV}, \cite{AVVVKS}.}
function of one complex variable
$x$
and of one real parameter
$y$.
In the system of layers
$$
B_{n-\frac{1}{2} <x< n+\frac{1}{2}}  \quad n=0,\pm1,...
$$
$A(x,y)$
has the analogous asymptotic behavior. The layers
$B_x$
and
$B_y$
intersect as it is shown in
Figure~\ref{Betabootpic}.

The residues of
$A(x,y)$
at poles in variable
$x$
are given by the expression
\begin{equation}
 r_n(y) \equiv Res {\bigr|}_{x=n+\frac{1}{2}} A(x,y) =
 \frac{1}{n!}( \textstyle \frac{1}{2}+y)\cdots (\frac{1}{2}+y+n) \equiv
 \displaystyle
 \frac{1}{n!} \textstyle \,(y+\frac{1}{2})_{(n+1)},
 \; \; n=0,1,... \,,
 \label{resX}
\end{equation}
were
$
(y+\frac{1}{2})_{(n+1)}
$
stands for the so-called Pochhammer symbol (shifted factorial).
Residues at poles in
$y$
read:
\begin{equation}
\rho_n(x) \equiv Res {\bigr|}_{y=n+\frac{1}{2}} A(x,y) =
\frac{1}{n!} \textstyle \,(x+\frac{1}{2})_{(n+1)},
\; \; n=0,1,... \,.
\label{resY}
\end{equation}

It is often said that the string amplitude is constructed solely
from the pole contributions of one channel:
$$
A(x,y)= \left[{{\text{Smooth}} \atop {\text{background}} } \right]
+
\sum_n \left[{{\text{Pole}} \atop {\text{contributions}} } \right].
$$
However in
\cite{AVVV}
it was shown that the meaning of this statement needs to be refined.
The technique of Cauchy forms known from the complex analysis allows
one to represent
$A(x,y)$
in the system of layers
$B_x, \, B_y$
as the uniformly converging series of the following structure:
$$
A(x,y)= \left[{{\text{Smooth}} \atop {\text{background}} } \right]
+
\sum_n \left[{{\text{Pole}} \atop {\text{contributions}} } +
{{\text{Correcting}} \atop {\text{polynomials}} }
 \right].
$$
This is not just a sum of poles in one channel, because the
cross-channel poles make explicit contributions to the correcting
polynomials as well as to background term. The degree of the
correcting polynomial depends on the asymptotic behavior in a given
layer.

For example, in the layer
$
B_{ -\frac{1}{2}<y<\frac{1}{2}}
$
the asymptotic behavior of
$A(x,y)$
corresponds to zero value of the bounding polynomial degree (the
amplitude grows slower than a linear function of
$x$
but faster than a constant). Hence in the Cauchy expansion we have to
take account of the correcting polynomials of
$0-$th
degree. Thus we obtain the following series:
\begin{equation}
A(x,y)=
\alpha_y(y)+
\sum_{n=0}^\infty
\frac{1}{n!}\,
\left(
\frac{(y+\frac{1}{2})_{(n+1)}}{x-n-\frac{1}{2}} +
\frac{(y+\frac{1}{2})_{(n+1)}}{n+\frac{1}{2}}
\right) ,\ \ \ \ \
y \in
\textstyle
(-\frac{1}{2}, \; \frac{1}{2}).
\label{CFormBy}
\end{equation}

In the layer
$B_{  -\frac{3}{2} <y< -\frac{1}{2}}$
$A(x,y)$
grows slower than a constant. So, we can write down the Cauchy
expansion without correcting polynomials:
\begin{equation}
A(x,y)=
\sum_{n=0}^\infty
\frac{r_{n}(y)}{(x-n-\frac{1}{2})}\ ,
\ \ \ \ \ \ \  y \in
\textstyle
(-\frac{3}{2},-\frac{1}{2}).
\label{CFormBy1}
\end{equation}
Note, that because the asymptotics becomes
``softer''
at large negative
$y$,
this expansion is also valid in every layer%
\footnote{Except the values corresponding to the coordinates
of ambiguity points.}
corresponding to
$y< -\frac{3}{2}$.
The similar Cauchy form can be written for
$A(x,y)$
in the layer
$B_{-\frac{3}{2}<x<-\frac{1}{2}}$:
\begin{equation}
A(x,y)=
\sum_{n=0}^\infty
\frac{\rho_{n}(x)}{(y-n-\frac{1}{2})}\ ,
\ \ \ \ \ \ \  x \in
\textstyle
(-\frac{3}{2},-\frac{1}{2}).
\label{CFormBx11}
\end{equation}

\section{Bootstrap for the String Amplitude}
\label{bootforSA}

\mbox

The full system of bootstrap equations for the string amplitude
contains three types of conditions:
\begin{itemize}
\item
The consistency conditions that arise naturally from the requirement
that the Cauchy form in one variable valid in certain layer can be
analytically continued into the perpendicular layer. For this reason
two Cauchy forms in different variables written in two intersecting
layers should coincide with one another in the intersection domain.

E.g., in the intersection domain
$B_{-\frac{5}{2}<y<-\frac{3}{2}}
\cap B_{-\frac{3}{2}<x<-\frac{1}{2}}$
(see
Fig.~{\ref{Betabootpic}})
we obtain the following condition:
\begin{equation}
 \Psi_{-1, -1}(x,y)=\sum_{n=0}^\infty
\frac{\rho_n(x)}{y-n-\frac{1}{2}}
-\sum_{n=0}^\infty
\frac{r_n(y)}{x-n-\frac{1}{2}}
\equiv
0, \ \  \; \;  x \sim -1, \; y \sim -2.
\label{VBoot}
\end{equation}
With the help of generating function
$\Psi_{-1,-1}(x,y)$
this condition can be rewritten as follows:
\begin{equation}
\frac{\partial^{k+p}}{\partial x^{k}\partial
y^{p}}\Psi_{-1,-1}(x,y){\Bigr|}_{x=-1 \atop y=-2}=0, \ \ \ \ \ \ \
\forall \  k,p=0,1,... \ .
\label{Vboot1}
\end{equation}

\item
The {\em collapsing conditions} for excessive degrees of the
correcting polynomials.

E.g., the Cauchy series
(\ref{CFormBy})
written for
$A(x,y)$
in the layer
$B_{-\frac{1}{2}<y<  \frac{1}{2}}$
is also valid in the lower layers with softer asymptotics. However in
those layers this expansion should coincide with
(\ref{CFormBy1})
and we obtain the corresponding collapsing condition:
$$
\alpha(y)= -\sum_{n=0}^\infty
 \frac{r_{n}(y)}{n+\frac{1}{2}}, \ \ \ y< \textstyle -\frac{1}{2}.
$$

\item
The
{\em superconvergency conditions}
in the layers with rapidly decreasing asymptotics.

E.g., in the layer
$B_{-\frac{7}{2}<y<-\frac{5}{2}}$
the string amplitude is the
($-2$)-bounded
function of the complex variable
$x$.
Thus one can write down the collapsing condition of the previous type
for the function
$xA(x,y)$
that is
$-1$-bounded
in this layer. It reads:
$$
\sum_{n=0}^{\infty}
r_{n}(y) =0, \ \ \ y< -\textstyle \frac{5}{2}.
$$
\end{itemize}

This system is by no doubt badly overdetermined and the problem of
pointing out its full subsystem rests a serious challenge. We will
return to this question in
Sec.~\ref{mmeth}.

\section{Numerical Tests}
\mbox

The numerical testing of sum rules for the resonance parameters is one
of the ways to check the consistency of our EFT approach. In this
Section we consider various situations that may encounter in realistic
EFT models by analyzing the numerical tests of sum rules following
from the above-described
``toy bootstrap model''.

To perform these numerical tests we first need to define a quantity
that would allow us to characterize the accuracy of saturation of a
given sum rule after taking into account the finite number of items.
Because of the symmetry of spectrum, it looks natural to
consider the difference of the contributions from the
$B_x$-layer
and
$B_y$-layer
poles at every step of the computation.

Usually the contributions of the first few poles have a definite (say,
positive) sign but, starting from a certain number
($N_{+}$),
the sign changes. Thus the saturation of a sum rule is provided by the
negative contribution of the long tail of distant poles. It
compensates gradually the positive contribution of the first few
items. As the convergence characteristics we chose the ratio:
$$
D(N) \equiv \frac{\Delta S(N)}{S_+} \cdot 100 \% ,
$$
where
$\Delta S(N)$
is  the discrepancy that remains after one considers
$N$
poles in the
$B_x$-layer
and
$N$
in
$B_y$-layer;
$S_+$
is the sum
{\em of all}
positive contributions (those corresponding to the finite number of
the first terms)%
\footnote{Certainly it only makes sense for
$N > N_+$.}.

As an example we perform the numerical testing of the sum rule that
follows from the bootstrap condition
(\ref{Vboot1})
with different
$p$
and
$k$.
Certain sum rules of this group are saturated very fast. For example,
for
$k=0, \, p=2$
first
$10$
poles saturate the sum rule with
$10 \%$
discrepancy. The sum rule with
$k=0, \, p=1$
saturates even faster: first
$10$
poles give
$3 \%$
discrepancy.

It is interesting to note that the scenario of saturation of this sum
rule is in accordance with the so-called
{\em local cancellation hypothesis}
(see
\cite{Schechter})
employed in phenomenological models. It says, that the properties of
the amplitude are mostly defined by the resonances with masses close
to the energy scale under consideration and, possibly, by nonsingular
background. The contributions from all the other resonances
``almost cancel among themselves''.
The fast saturating sum rules provide us with powerful constrains for
the parameters of first few poles. The hope is that the reliable sum
rules which were found for
$\pi N, \pi K$
and
$KN$
systems are exactly of this type and, hence, they can be used to study
the hadron spectrum.

\begin{figure}[thb]
\centerline{\psfig{figure=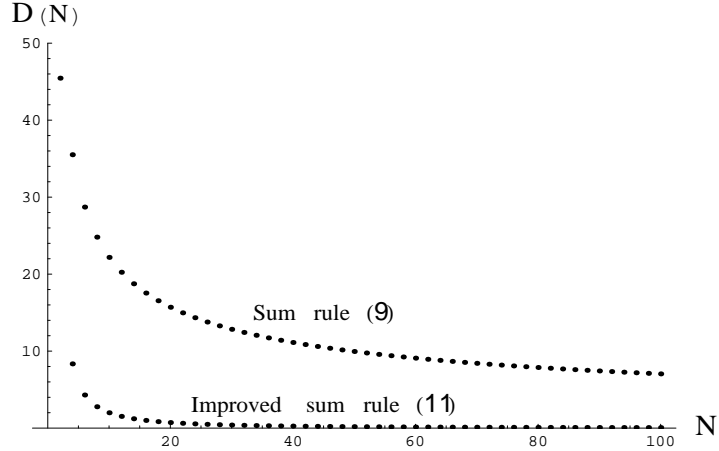,height=6cm,width=10cm}}
\caption{Saturation of sum rule
(\ref{VbootWCP}) and of improved sum rule (\ref{VbootWDer}).
\label{NT1} }
\end{figure}

However, we will show that the fast saturation is not always the case
and in certain situations the important properties of the amplitude
may depend on the large number of distant poles. For this let us consider the
sum rule that follows from
(\ref{Vboot1})
with
$k=0, \, p=0$:
\begin{equation}
\Psi_{-1,-1}(x,y)\big|_{x=-1 \atop y=-2}=0.
\label{VbootWCP}
\end{equation}
The convergence of this sum rule turns out to be very slow. The
dependence of the relative discrepancy
$D$
on the number of poles taken into account is shown in
Fig.~\ref{NT1}.

To saturate this sum rule with high accuracy on has to use the
detailed information about the resonance spectrum (poles and
residues). For example, to saturate it with
$99\%$
 accuracy
($1\%$
discrepancy)
one has to take into account more than
$5000$
poles.

Now we would like to discuss a way that allows one to modify this sum
rule so that it will be possible to saturated it with smaller number
of poles. The trick we are using is very similar to the conventional
method of accelerating the convergency: the partial summation of
converging series. Instead of
(\ref{CFormBy1})
and
(\ref{CFormBx11})
representing
$A(x,y)$
in
$
B_{\frac{5}{2}<y< \frac{3}{2}}
$
and
$
B_{-\frac{3}{2}<x<-\frac{1}{2}}
$
we now make use the Cauchy forms with excessive degree of the correcting
polynomial.

The bootstrap conditions in
$
B_{-\frac{5}{2}<y<-\frac{3}{2}} \cap B_{-\frac{3}{2}<x<-\frac{1}{2}}
$
then read:
\begin{equation}
\begin{split}
&\alpha_x(x)-\alpha_y(y)+  {\Psi}_{0,0}(x,y)= \\&
\alpha_x(x)-\alpha_y(y)+\sum_{n=0}^\infty
 \left(
\frac{\rho_n(x)}{y-n-\frac{1}{2}}+
\frac{\rho_n(x)}{n+\frac{1}{2}}
\right)
-\sum_{n=0}^\infty
\left(
\frac{r_n(y)}{x-n-\frac{1}{2}}+
 \frac{r_n(y)}{n+\frac{1}{2}}
\right)
\equiv
0, \ \    x \sim -1, \; y \sim -2.
\end{split}
\end{equation}

We consider the sum rule:
\begin{equation}
\alpha_x(x)|_{x=-1}-\alpha_y(y)|_{y=-2}+
 {\Psi}_{0,0}(x,y)\big|_{x=-1 \atop y=-2}=0.
\label{VbootWDer}
\end{equation}

The  smooth parts of the amplitude (background terms)
$\alpha_x(x)$
and
$\alpha_y(y)$
can be found, for example, from the bootstrap conditions in another
layers. From
$
B_{-\frac{5}{2}<x<-\frac{3}{2}} \cap B_{-\frac{5}{2}<y<-\frac{3}{2}}
$
(using
(\ref{CFormBy}), (\ref{CFormBx11}))
we find:
\begin{equation}
\alpha_y(y)= \left[
\sum_{n=0}^\infty
\frac{\rho_{n}(x)}{(y-n-\frac{1}{2})}-
\sum_{n=0}^\infty
\left(
\frac{r_n(y)}{x-n-\frac{1}{2}} +
\frac{r_n(y)}{n+\frac{1}{2}}
\right)
\right]_{x=-2}.
\end{equation}
In the similar way from
$
B_{-\frac{3}{2}<x<-\frac{1}{2}} \cap B_{-\frac{7}{2}<y<-\frac{5}{2}}
$:
\begin{equation}
\alpha_x(x)= \left[
\sum_{n=0}^\infty
\frac{r_{n}(y)}{(x-n-\frac{1}{2})}-
\sum_{n=0}^\infty
\left(
\frac{\rho_n(x)}{y-n-\frac{1}{2}} +
\frac{\rho_n(x)}{n+\frac{1}{2}}
\right)
\right]_{y=-3}.
\end{equation}

The sum rule
(\ref{VbootWDer})
saturates very fast: taking into account of just
$10$
first poles one obtains
$98\%$
accuracy. The contribution of distant poles (in both variables) is
negligible. So, in principle, from this sum rule we also can obtain
the constrains for the values of resonance parameters of few first
poles.

The studies of possibility to apply the methods of convergency
acceleration for the case of realistic EFT models describing hadron
scattering to cure certain
``bad''
sum rules and to derive the fastly saturated sum rules  are now in
progress.


\section{Matrix Methods}
\label{mmeth}
\mbox

The problem of analysis of the system of bootstrap constrains is
closely linked to the general theory of analytic continuation. In
fact, the bootstrap system expresses the possibility to carry out the
analytic continuation of the meromorphic function%
\footnote{This only relates to the lowest order (tree  level)
constraints. When considering the higher level systems one deals with
another kinds of singular points.}
with given asymptotic regime from one layer to the perpendicular one.
Bootstrap constrains result in a set of restriction for the resonance
parameters of a theory. One of the most important (and difficult)
problems appearing in our EFT approach is  formulated as  follows:
``Should we take into account the bootstrap constraints in all the
intersection domains of various layers?  Do the bootstrap conditions
in the intersection domains of other layers impose any new restrictions
on the parameters of the theory in addition to those following from the
bootstrap system in a given intersection domain, or they are just a
consequence of the bootstrap in one domain?''.
To study this and the related problems we introduce matrix methods
that are of great use in general theory of analytic continuation.

Let us consider the analytic continuation of the analytic (in certain
domain
$D$)
function of two complex variables. The coefficients
$a_{ij}$
of the double Taylor series in the vicinity of
$(x_1,y_1)$:
$$
F(x,y)=
\sum_{i,j}a_{ij}(x_1,y_1) (x-x_1)^i(y-y_1)^j \ \ \
x \sim x_1 \ \ y \sim y_1
$$
are linked to the coefficients
$\tilde{a}_{ij}$
of the expansion around
($x_2,y_2$)
by way of certain matrix operation
$\hat{T}(x_2,y_2; x_1,y_1)$:
$$
\tilde{a}_{ij}(x_2,y_2)=\sum_{k,l}
\hat{T}_{ijkl} (x_2,y_2; x_1,y_1) a_{kl}(x_1,y_1).
$$
In fact, the matrix
$\hat{T}$
contains the complete information about non-trivial analytic
properties of the function
$F$.
On the other hand it is this matrix that transforms bootstrap
conditions written in the intersection domain of two given layers into
the bootstrap system in other domains.

For string amplitude it is possible to present the specific  matrix of
analytic continuation from, say,
$(x_1,y_1)$
to
$(x_1+1,y_1)$.
This matrix mirrors the recurrent properties of the
$B$-function
and transforms the bootstrap system at
$(x_1,y_1)$
to another bootstrap system at
$(x_1+1,y_1)$:
\begin{equation}
\tilde{a}_{ij}(x_1+1,y_1) = \sum_{k=0 }^{i }\sum_{l=0 }^{j }
\hat{T}_{ijkl} (x_1 + 1,y_1; x_1,y_1) a_{kl}(x_1,y_1)\ ,
\label{matraction}
\end{equation}
where
\begin{equation}
\hat{T}_{ijkl} (x_1 + 1,y_1; x_1,y_1) \equiv
\left.
   \frac{1}{(i-k)!(j-l)!}
\left[
 \frac{\partial^{i+j-k-l}}{\partial x^{i-k}\partial y^{j-l}}
 \left( \frac{x +y +1}{x +\frac{1}{2}}
 \right)
\right]
\right|_{(x_1,y_1)}\, ;
\label{matrix}
\end{equation}
$a_{ij}$
and
$\tilde{a}_{kl}$
stand for the sets of Taylor expansion coefficients in the
vicinity of
$(x_1,y_1)$
and
$(x_1+1,y_1)$
respectively. The bootstrap system requires that in the domain of
intersection of layers the set of Taylor coefficients calculated from
the Cauchy expansion in one layer should coincide with that calculated
from the Cauchy expansion in the perpendicular layer. So it is exactly
the matrix
(\ref{matrix})
acting as shown in
(\ref{matraction})
that allows one to pass from the bootstrap system in a given domain to
the system in the neighboring domains.

Employing the information on the asymptotic behavior of the string
amplitude that is contained in collapsing and superconvergence
conditions (see
Sec.~\ref{bootforSA}),
it turns out possible to show that -- due to recurrent properties of
the string amplitude -- the bootstrap systems appearing in all the
intersection domains of various layers are, in fact, equivalent to one
another.

It is possible that the application of methods of the general theory
of analytic continuation and of the technique of infinite dimensional
matrices for the case of realistic EFT models would also help to
single out the independent subsystem of the full system of bootstrap
constraints.

\section{Conclusions}
\mbox

The exactly solvable
``toy bootstrap model''
based on Veneziano string amplitude is an outstanding proving ground
for the development of methods to handle the bootstrap systems. The
hope is that the results we obtained in the framework of this model
will deepen our understanding of analytical properties of scattering
amplitudes in the EFT approach.

\section*{Acknowledgements}
\mbox

This work was supported by INTAS (project 587, 2000) and by Ministry
of Education of Russia (Programme ``Universities of Russia''). I am
grateful to A.~Vereshagin and V.~Vereshagin for multiple help and
advise and to M.~Vyazovski for stimulating discussions.


\end{document}